\documentstyle[12pt]{article}
\input epsf

\begin{document}

\begin{titlepage}

\begin{flushright}
Freiburg--THEP 97/19\\
August 1997
\end{flushright}
\vspace{2.5cm}

\begin{center}
\large\bf
{\LARGE\bf Nonperturbative Higgs propagator: 
           NLO correction in the $1/N$ expansion}\\[2cm]
\rm
{A. Ghinculov\footnote{Address after October 1, 1997: 
                       Randall Laboratory of Physics, University of Michigan, 
		       Ann Arbor, Michigan 48109--1120, USA}
, T. Binoth and J.J. van der Bij}\\[.5cm]

{\em Albert--Ludwigs--Universit\"{a}t Freiburg,
           Fakult\"{a}t f\"{u}r Physik}\\
      {\em Hermann--Herder Str.3, D-79104 Freiburg, Germany}\\[3.5cm]
      
\end{center}
\normalsize

\begin{abstract}
We derive a nonperturbative result for the Higgs propagator
by calculating the NLO correction 
in the $1/N$ expansion of $O(N)$ sigma models.
For dealing with the $1/N$ expansion beyond leading order,
we develop techniques to calculate certain infinite sets of multiloop 
Feynman diagrams. 
We then calculate the Higgs lineshape
in fermion scattering, as seen for instance at a muon collider. 
We compare with the existing two--loop perturbative results. 
There is a considerable
discrepancy between perturbation theory and the $1/N$ expansion in LO.
By including the NLO correction, this discrepancy is reduced dramatically.
The results are in very good agreement with the two--loop 
result even for masses as high as 850 GeV. A maximum of the mass
is reached at 930 GeV. 
For phenomenological purposes, we give a simple approximate 
relation between the Higgs mass and width.
\end{abstract}


\end{titlepage}


\title{Nonperturbative Higgs propagator: 
       NLO correction in the $1/N$ expansion}

\author{A. Ghinculov\thanks{Address after October 1, 1997: 
                       Randall Laboratory of Physics, University of Michigan, 
		       Ann Arbor, Michigan 48109--1120, USA}
, T. Binoth and J.J. van der Bij}

\date{{\em Albert--Ludwigs--Universit\"{a}t Freiburg,
           Fakult\"{a}t f\"{u}r Physik},\\
      {\em Hermann--Herder Str.3, D-79104 Freiburg, Germany}}

\maketitle

\begin{abstract}
We derive a nonperturbative result for the Higgs propagator
by calculating the NLO correction 
in the $1/N$ expansion of $O(N)$ sigma models.
For dealing with the $1/N$ expansion beyond leading order,
we develop techniques to calculate certain infinite sets of multiloop 
Feynman diagrams. 
We then calculate the Higgs lineshape
in fermion scattering, as seen for instance at a muon collider. 
We compare with the existing two--loop perturbative results. 
There is a considerable
discrepancy between perturbation theory and the $1/N$ expansion in LO.
By including the NLO correction, this discrepancy is reduced dramatically.
The results are in very good agreement with the two--loop 
result even for masses as high as 850 GeV. A maximum of the mass
is reached at 930 GeV. 
For phenomenological purposes, we give a simple approximate 
relation between the Higgs mass and width.
\end{abstract}


\section{Introduction}

With the discovery of the vector bosons and of the top quark behind us, 
the next major question
in the phenomenology of the weak interactions is the search 
for the Higgs boson. Within the standard model the properties of the Higgs boson 
are determined when its mass is fixed, as the tree level selfcouplings 
are proportional to $m_H^2$. As long as the Higgs particle is light,
there are no fundamental problems in determining its properties 
from perturbation theory. 
However, when the Higgs particle becomes heavy, ${\cal O}(TeV)$,
the selfcoupling becomes large, and perturbation theory becomes unreliable.

Therefore one would like to find an approximation to Higgs physics 
beyond perturbation theory which
is applicable for large couplings as well. When one realizes 
that the Higgs sector of the standard model is nothing but an 
$O(4)$ linear sigma model, the
expansion in $1/N$ of the $O(N)$-symmetric sigma model suggests itself.

The $1/N$ expansion has a long history \cite{einhorn}---\cite{coleman}, 
and has also been applied
to the Higgs boson of the standard model \cite{einhorn,casalbuoni}. 
The results in 
leading order in $1/N$ are interesting. They indicate a saturation 
of the Higgs boson mass when the coupling increases. Also the relation 
between the Higgs width and mass is very different at larger couplings. 
However, the situation at leading order 
is not satisfactory because the lowest order 
is numerically very far from perturbation theory for small Higgs mass. 
Perturbation theory has recently even been extended
to the two-loop level \cite{2loop,riess,jikia}, and should be well
under control for small Higgs mass.
This shed for a long time doubts on the relevance of the $1/N$ approach.
In order to clarify the situation it is therefore clearly desirable 
to know the next--to--leading order contribution in $1/N$. 

What makes the $1/N$ expansion particularly interesting 
is that it has certain features which are absent 
in perturbation theory, and which are to be expected 
from an exact solution. One feature is the absence in physical results
of any renormalization scheme dependence. As we shall see, the 
next--to--leading order solution which we will derive 
has another such feature, namely the finiteness
of wave function renormalization constants. Interestingly enough,
this occurs after the summation of Feynman diagrams which are all
ultraviolet divergent.

The $O(N)$ sigma model at leading order in the $1/N$ expansion was 
discussed in detail by a number of authors. These works study 
the effective potential and the question of the occurrence of spontaneous 
symmetry breaking in $O(N)$ models, as well as two-- and four--point Green
functions. These aspects were studied from two points of view. One attempts
to treat the $O(N)$ sigma model as a fundamental, renormalized theory; 
and the other as an effective theory, with a built--in cutoff.
Although a large amount of work was done, most of it is limited 
to the leading order.
Root \cite{root} wrote down expressions for the 
effective potential in higher order, 
but did not evaluate them explicitly. The reason for this 
is twofold. First of all, the expressions are quite complicated. 
More fundamental is the occurrence of a 
tachyon in the theory.

A striking feature of the leading order solution is the presence
in the propagators of a tachyonic pole. As long as one is concerned 
only with the leading order, and the Higgs coupling is not very
strong, the causality violating effects induced by the tachyon 
are at least numerically negligible. They are suppressed essentially 
by the Landau scale. However, if one wants to calculate higher
order corrections, these effects are unavoidable because the tachyon
appears then in loops and leads to pathological solutions. For instance,
it was shown by Root \cite{root} that the effective potential becomes
complex in next--to--leading order. 

The tachyon problem is similar to the Landau ghost in perturbation theory. 
However, the occurrence and position of the Landau ghost relies on
using a perturbative expression precisely where one knows this expression
is not a good approximation anymore. For the case of the tachyon
problem in the $1/N$ expansion this is no more the case.
The propagators in the $1/N$ expansion are
summed to all orders in the coupling constant. Thus their validity does not
depend on the strength of the coupling.
A possible interpretation for the presence of the tachyon is that it
indicates the triviality of the theory. 

One simple and consistent way to deal with the tachyon 
is to regard the $O(N)$ model as
an effective theory and to introduce explicitly a cutoff under 
the tachyon scale \cite{nunes}. However, we prefer to use in this paper 
a different approach. We introduce a scheme to subtract minimally
the tachyon from the Green functions order by order in the $1/N$ expansion.
This procedure is well--adapted to the calculation of
higher order effects in $1/N$.

The outline of this paper is the following. In section 2 we 
introduce the model and the leading order for fixing the notations. 
In the following section we discuss and motivate in detail the prescription for 
the treatment of the tachyon problem. In section 4 we describe 
the next--to--leading order calculation and the methods for evaluating 
the multiloop diagrams involved. In section 5 we treat the Higgs 
lineshape nonperturbatively and discuss the results. Section 6 summarizes
the conclusions of the paper.


\section{Renormalization and leading order solution}

The $O(N)$ sigma model at leading order in the $1/N$ expansion was already
discussed extensively in the literature. 
For this reason, we will not repeat here this discussion. In the following,
for fixing the notations, we will only derive the two--point functions, 
assuming that the theory has a ground state with spontaneously broken symmetry.

The starting point of the calculation is the following Lagrangian:

\begin{equation}
  {\cal L} =   \frac{1}{2}            \partial_{\nu}\Phi_0 \partial^{\nu}\Phi_0 
             - \frac{\mu_0^2}{2}      \Phi_0^2 
	     - \frac{\lambda_0}{4! N} \Phi_0^4 
\end{equation}
                          
Here $\Phi$ is a scalar field with $N$ components $\phi^i$, $i=1, ...,N$.
In principle one could use this Lagrangian in order to calculate 
contributions in different orders of 1/N. However, the combinatorics becomes
very complicated and it is advantageous to introduce an auxiliary field $\chi$,
and add a nondynamical term to the Lagrangian according to a combinatorial 
trick proposed in ref. \cite{coleman}:

\begin{eqnarray}
  {\cal L} & = & {\cal L} + \frac {3 N}{2 \lambda_0} 
                             (\chi_0 - \frac{\lambda_0}{6 N} \Phi_0^2 - \mu_0^2)^2 
	\nonumber \\	
           & = &				     
    \frac{1}{2} \partial_{\nu}\Phi_0 \partial^{\nu}\Phi_0 
  - \frac{1}{2} \chi_0 \Phi_0^2 
  + \frac{3 N}{2 \lambda_0} \chi_0^2
  - \frac{3 \mu_0^2 N}{\lambda_0} \chi_0 + const. 
\end{eqnarray}
 
This form of the Lagrangian has the same physical content as 
before. However, the Feynman rules are changed, thereby facilitating 
the counting of powers of $1/N$. 
For doing higher order calculations, the introduction of the extra
field $\chi$ is practically the only way not to get lost in the 
combinatorics -- see also ref. \cite{root}.

The fields and coupling constants in eqns. 1, 2 are bare quantities.
In order to perform renormalization, we make the following substitutions:

\begin{eqnarray}
  \frac{3}{\lambda_0}       &=&   \frac{3}{\lambda} + \Delta \lambda    \\
  \frac{3 \mu_0}{\lambda_0} &=& - \frac{v^2}{2} (1 + \Delta t_{\sigma}) \\
  \phi_i^0 &=&  \pi_i  Z_{\pi} ~~~ , ~~~~  i=1,...,N-1                  \\
  \phi_N^0 &=&  \sigma Z_{\sigma} + \sqrt{N} v                          \\
  \chi^0   &=&  \chi   Z_{\chi} + \hat \chi +\Delta t_{\chi}
\end{eqnarray}

Here $\sqrt{N} v$ and $\hat{\chi}$ are 
the expectation values of the fields
$\Phi$ and$\chi$ in the broken symmetry ground state.
For the case of the standard model Higgs sector, $N = 4$ and $v=123$ GeV.
$\Delta t_{\sigma}$ and $\Delta t_{\chi}$ are the tadpole counterterms 
corresponding to the fields $\sigma$ and $\chi$.

The tadpole counterterms $\Delta t_{\sigma}$ and $\Delta t_{\chi}$ can be determined 
in the following way. In the original Lagrangian of eq. 1 only one tadpole
counterterm is needed. Upon introduction of the auxiliary field $\chi$
two such counterterms are present, but they are related to each other.
Some contributions can be shifted from the definition 
of one counterterm to the other.
It is then convenient for performing a next--to--leading 
order calculation to define 
the tadpole counterterms so that the vacuum expectation 
values of the fields $\sigma$
and $\chi$ do not receive corrections in higher orders. 
By solving the gap equation 
in leading order, it is well--known that one finds in the broken symmetry 
ground state $\hat{\chi} = 0$. By
requesting that this simple relation be preserved in higher orders, 
we fix uniquely the 
relation between the two tadpole counterterms 
$\Delta t_{\sigma}$ and $\Delta t_{\chi}$.
Therefore, we define $\Delta t_{\sigma}$ and 
$\Delta t_{\chi}$ by the condition
that  the one-point functions for the fields $\sigma$ and $\chi$ vanish. 
It is easy to convince oneself that this condition ensures automatically 
that the Goldstone theorem is valid, and the $\pi$ fields remain massless.

Within perturbation theory, the counterterms defined above are 
power series of the renormalized coupling constant $\lambda$. In the 
$1/N$ expansion they are power series in $1/N$. 
As it turns out, the contributions to most of these counterterms vanish
in the leading order of the $1/N$ expansion:

\begin{eqnarray}
  \Delta \lambda    &=&     \delta \lambda^{(0)} + \frac{1}{N} \delta \lambda 
                                                          + {\cal O}\left(\frac{1}{N^2}\right) \nonumber \\
  Z_{\pi}           &=& 1 + \frac{1}{N} \delta Z_{\pi}    + {\cal O}\left(\frac{1}{N^2}\right) \nonumber \\
  Z_{\sigma}        &=& 1 + \frac{1}{N} \delta Z_{\sigma} + {\cal O}\left(\frac{1}{N^2}\right) \nonumber \\
  Z_{\chi}          &=& 1 + \frac{1}{N} \delta Z_{\chi}   + {\cal O}\left(\frac{1}{N^2}\right) \nonumber \\
  \Delta t_{\sigma} &=&  \frac{1}{N} \delta t_{\sigma}    + {\cal O}\left(\frac{1}{N^2}\right) \nonumber \\
  \Delta t_{\chi}   &=&  \frac{v^2}{N} \delta t_{\chi}      + {\cal O}\left(\frac{1}{N^2}\right)
\end{eqnarray}

\begin{figure}
\begin{tabular}{ccccc}
 \hspace{2cm}
 $i (N-1) \alpha^{(0)}(s)$ & $=$ & 
                  \raisebox{-.74cm}{\epsfxsize = 3.3cm \epsffile{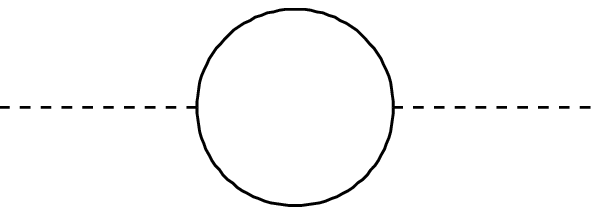}} & + & counterterm 
\end{tabular}
\caption{{\em The leading order bubble diagram.}}
\end{figure}

The only counterterm which is needed at leading order in $1/N$
is actually the $\delta \lambda^{(0)}$ counterterm 
of the quartic interaction.

In order to calculate the propagators, one has to evaluate the 
$\sigma \sigma$, $\chi \sigma$, $\chi \chi$ and $\pi \pi$ self--energy diagrams, 
and then to invert the matrix.	
As it turns out, the only nontrivial contribution to the proper self--energy
functions of the theory in leading order of the $1/N$ expansion is due to 
the diagram shown in fig. 1. The ultraviolet divergence of the bubble integral
is absorbed in the coupling constant renormalization, $\delta \lambda^{(0)}$.
One finds the following well--known leading order propagators:

\begin{eqnarray}
  D_{\sigma \sigma}(s) &=&                        \frac{i         }{s - m^2(s)} \nonumber \\
  D_{\chi \chi}(s)     &=& \frac{1}{      N  v^2} \frac{i s m^2(s)}{s - m^2(s)} \nonumber \\
  D_{\chi \sigma}(s)   &=& \frac{1}{\sqrt{N} v  } \frac{i   m^2(s)}{s - m^2(s)} \nonumber \\
  D_{\pi^i \pi^j}(s)   &=& \delta_{ij}            \frac{i}{s}
\end{eqnarray}

Here, the quantity $m^2(s)$ plays the r\^ole of an energy dependent mass, and
is given by the bubble diagram of fig. 1:

\begin{equation}
  m^2(s) = \frac {v^2}{\frac{3}{\lambda} + \alpha^{(0)}(s)} 
    ~~~~~  \equiv  ~~~~~ 
            \frac {v^2}{\frac{3}{\lambda} - \frac{1}{32 \pi^2} 
                        \log{\left(-\frac{s+i \epsilon}{\mu^2}\right)}} 
\end{equation}
where $\mu$ is the subtraction scale of the bubble diagram. 

We would like
to reemphasize that, although a renormalization scale appears in eqns. 9,
the physical predictions of this solution are  
free of any residual renormalization scheme dependence. 
This is because perturbation theory was summed up in all orders.
Whenever one uses this
solution of the $O(N)$ sigma model to relate physical observable quantities, 
the renormalization scale $\mu$ can be eliminated from the result. 
As an example, one can convince oneself
that the relation between the Higgs mass and width is independent of the 
intermediary renormalization scheme used \cite{einhorn}.


\section{Tachyonic regularization}

As expected, the propagators $D_{\sigma \sigma}$, $D_{\chi \chi}$ and 
$D_{\chi \sigma}$ derived in the previous section
contain a pole corresponding to the Higgs boson. 
Apart from this, there is of course also a tachyonic 
pole in these expressions. 
Its Euclidian position, $s=-\Lambda_t^2$, is given by the following equation:

\begin{equation}
    \frac{v^2}{\Lambda_t^2} 
  - \frac{1}{32 \pi^2} \log{\left( \frac{\Lambda_t^2}{\mu^2} \right)}
  + \frac{3}{\lambda} = 0   ~~~~~ ,
\end{equation}
and differs from the position of the Landau pole 
$\Lambda_L = \mu e^{48 \pi^2/\lambda}$ by a correction of order
$v^2/\Lambda_L^2$.
The occurrence of this Euclidian pole creates problems because it 
spoils the causality of the theory.

Let us analyse the way one performs a calculation in the $1/N$ expansion.
To calculate, let's say, the Higgs propagator,
one first performs a double expansion of the self--energy in 
the coupling constant $\lambda$ and in $1/N$. 
Then each coefficient in the $1/N$ expansion is given by a power series 
in the coupling constant $\lambda$. For one given $1/N$ order,
due to the combinatorial structure,
one is able to calculate explicitly all Feynman diagrams 
in the perturbative expansion, and to sum up the $\lambda$ series. The result
then appears to contain an additional tachyonic pole. 
However, let us notice that the coefficient of the $1/N$ expansion which
we calculate is not defined uniquely by the expansion in 
$\lambda$ which we calculate by Feynman diagrams. 
One still has the freedom to add
a function in $\lambda$ whose perturbative expansion vanishes.
Our treatment of the tachyonic pole is based on the observation 
that its residuum is indeed such a nonperturbative function
of the coupling constant, of the type $e^{1/\lambda}$. One can
add an opposite contribution, so that the tachyonic pole is cancelled,
because this contribution vanishes exactly in all orders of 
perturbation theory. 

For this reason, we simply subtract the tachyon minimally at its pole from
the leading order propagators. With this prescription, the calculation
becomes well--defined, and one can represent the $O(1/N)$ contribution 
in terms of a number of Feynman--like graphs. Such a 
tachyonic regularization can be carried out order by order 
in the $1/N$ expansion.

Along the lines of the discussion above, we choose to regularize 
the propagators of eqns. 9 by subtracting the tachyon pole minimally. 
We introduce the following subtractions:

\begin{eqnarray}
  D_{\sigma \sigma}(s) &=& i \left(   \frac{1     }{s - m^2(s)}
                                 - \frac{\kappa}{s + \Lambda_t^2} \right) \nonumber \\
  D_{\chi \chi}(s)     &=& \frac{i s m^2(s)}{N  v^2} 
                          \left(   \frac{1     }{s - m^2(s)}
                                 - \frac{\kappa}{s + \Lambda_t^2} \right) \nonumber \\
  D_{\chi \sigma}(s)   &=& \frac{i m^2(s)}{\sqrt{N} v}  
                          \left(   \frac{1     }{s - m^2(s)}
                                 - \frac{\kappa}{s + \Lambda_t^2} \right)
\end{eqnarray}
were 

\begin{equation}
  \kappa =  \frac{1}{ 1 + \frac{\Lambda_t^2}{32 \pi^2 v^2} }
\end{equation}
is the residuum of the tachyonic pole in the $\sigma \sigma$ propagator. 

It is worthwhile noting in this context that the tachyon not only has a mass
of the order of the Landau pole mass. Also its residuum, {\em i.e.}
its coupling to the other fields, is suppressed essentially 
by the same mass scale. Therefore the effects of the tachyon 
on the low energy
physics are completely negligible as long as the mass of the 
Landau pole is not very low.

Let us examine the tachyonic regularization of eqns. 12.
When we use the tachyonic regularized expressions 
instead of eqns. 9
for calculating higher order Green function, we essentially modify
the Green function
by a quantity proportional to $\kappa$. As explained, 
this is a contribution which exactly vanishes
in all orders of perturbation theory because $\kappa$ is a function of the
type $e^{1/\lambda}$. 

Our tachyonic regularization can be seen
as a different prescription for summing up the perturbative 
expansion of a given
term in the $1/N$ expansion. This prescription is such that it does not
result in the presence of a tachyon in the theory's spectrum, which would
violate causality. At the same time, 
the original information we started with, that is, the power series
in $\lambda$ of the coefficient of the $1/N$ expansion, remains
untouched. No doubt, this prescription is not unique. For a trivial
theory, this ambiguity can be used for modeling the nondecoupling effects
associated with unknown physics at a higher energy scale, very much
in the same way a cutoff would do.
One expects this uncertainty to become important numerically 
only when the energy of the process
under consideration approaches the Landau scale.

Finally, we would like to stress that we do not interpret the
tachyonic regularization as a proof that the $1/N$ expansion is
free of tachyons. It only says that within the usual derivation
of the $1/N$ expansion the occurrence of tachyons is arbitrary.
To establish conclusively the presence or the absence 
of tachyons, one needs a procedure for calculating the coefficients
of the $1/N$ expansion without using Feynman diagrams at intermediary 
stages. Such a procedure is not available so far. Until substantial 
progress is made
in this direction, we prefer a scheme which respects causality and allows
one to perform higher order calculations consistently.


\section{Next-to-leading order calculation}

With the tachyonic regularized propagators defined at leading order in the
previous section, it is now possible to give a diagrammatic 
representation of the next-to-leading order contributions
to the self--energy functions. The graphs are given in fig. 2.
Each graph is in fact an infinite sum of multiloop Feynman diagrams which
all are of the same order in the expansion parameter $1/N$. 
This is shown explicitly in fig. 3 for one graph.

\begin{figure}[t]
\begin{tabular}{cccccccc}
      $i \alpha(s)$
    & $=$
    & 
    &       \raisebox{-.74cm}{$A_1$ \hspace{-1cm} \epsfxsize = 3.3cm \epsffile{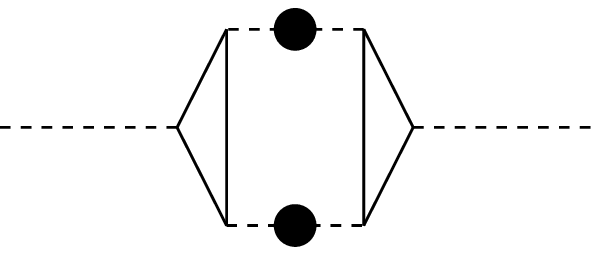} }  
    & $+$ & \raisebox{-.74cm}{$A_2$ \hspace{-1cm} \epsfxsize = 3.3cm \epsffile{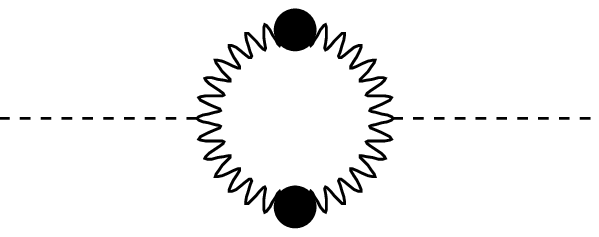} }  
    & $+$ & \raisebox{-.74cm}{$A_3$ \hspace{-1cm} \epsfxsize = 3.3cm \epsffile{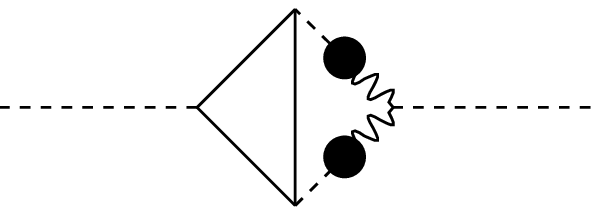} } 
  \\
    &  
    & $+$ & \raisebox{-.74cm}{$A_4$ \hspace{-1cm} \epsfxsize = 3.3cm \epsffile{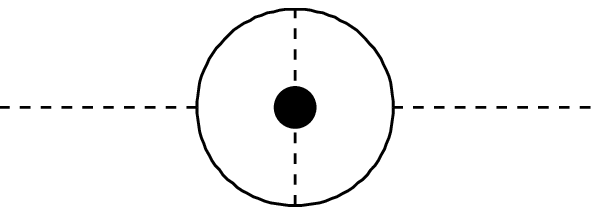} }  
    & $+$ & \raisebox{-.74cm}{$A_5$ \hspace{-1cm} \epsfxsize = 3.3cm \epsffile{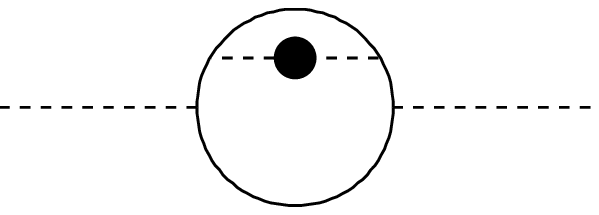} }
    & $+$ & counterterms
  \\
      $i \frac{v^2}{N} \beta(s)$
    & $=$
    & 
    &       \raisebox{-.74cm}{$B_1$ \hspace{-1cm} \epsfxsize = 3.3cm \epsffile{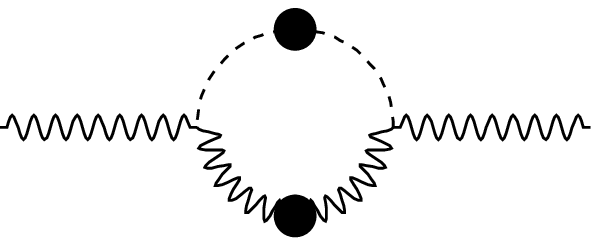} }  
    & $+$ & \raisebox{-.74cm}{$B_2$ \hspace{-1cm} \epsfxsize = 3.3cm \epsffile{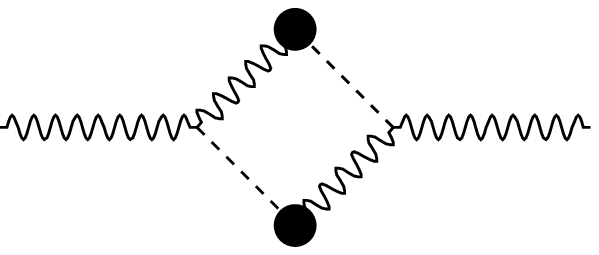} } 
    &  $+$ & counterterms
  \\
      $i \frac{v}{\sqrt{N}} \gamma(s)$
    & $=$
    & 
    &       \raisebox{-.74cm}{$C_1$ \hspace{-1cm} \epsfxsize = 3.3cm \epsffile{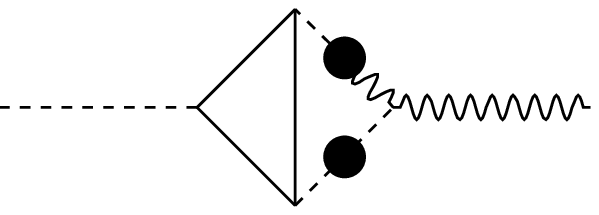} }  
    & $+$ & \raisebox{-.74cm}{$C_2$ \hspace{-1cm} \epsfxsize = 3.3cm \epsffile{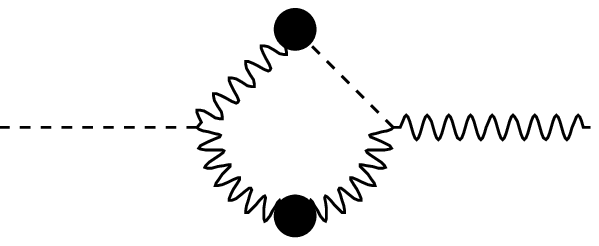} } 
    & $+$ & counterterms
  \\
      $i \frac{v^2}{N} \delta(s)$
    & $=$
    & 
    & \raisebox{-.74cm}{$D$ \hspace{-1cm} \epsfxsize = 3.3cm \epsffile{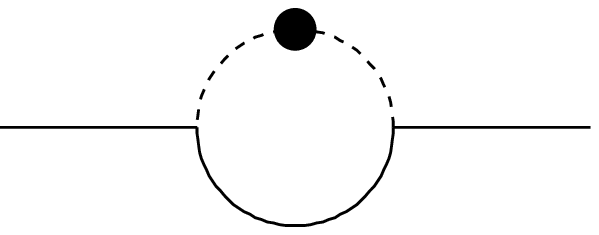} }
    &  $+$ & counterterms
    & 
    & \raisebox{-.74cm}{\epsfxsize = 3.3cm \epsffile{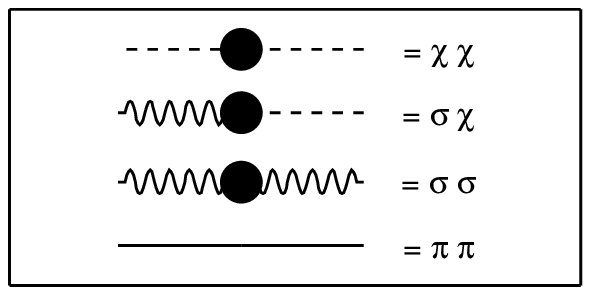}}
\end{tabular}
\caption{{\em Infinite sums of multiloop Feynman diagrams which contribute
              in next--to--leading order in $1/N$
              to the two--point functions of the $O(N)$ sigma model.
	      The blob on propagators denotes the summed--up leading order
	      propagators. Note that the $\pi \pi$ propagator at leading
	      order in $1/N$ is a free propagator.
	      One of the graphs above is shown in expanded form in fig. 3.}}
\end{figure}

With the next--to--leading order self--energy functions defined
in fig. 2, it is straightforward to calculate the propagators.
In particular, the $\sigma \sigma$ propagator, which is
our main concern in this paper, is given by:

\begin{equation}
  D^{(NLO)}_{\sigma \sigma}(s) = \frac{i}{s - m^{2 ~ (corr)}(s)} ~~~~~,
\end{equation}
where

\begin{equation}
  m^{2 ~ (corr)}(s) = \frac {v^2}{ \frac{3}{\lambda} + \alpha^{(0)}(s)
     + \frac{1}{N} \left\{  -  \alpha^{(0)}(s)        +    \alpha(s) 
                     + 2       \frac{v^2}{m^2(s)}          \gamma(s) 
                     +   \left[\frac{v^2}{m^2(s)}\right]^2 \beta(s) 
                    \right\}      } 
\end{equation}

Because of the complicated form of the expressions, 
the evaluation of these graphs
is highly involved and can only be performed numerically. 
In order to do the actual calculation new techniques were necessary.
We developed techniques which extend the methods of ref. \cite{3loop}
which deals with massive three--loop diagrams.
Some graphs are also ultraviolet divergent. It is rather complicated
to disentangle the poles in $\epsilon = n-4$ of the sets of multiloop diagrams
of the type shown in fig. 2. Therefore in order to perform renormalization
one first determines the counterterms in terms of multiloop graphs
by using the renormalization conditions discussed in the previous section.
Then, without evaluating the counterterms explicitly, 
one adds them to the actual graphs which contribute to
the self--energies $\alpha(s)$, $\beta(s)$, $\gamma(s)$ and $\delta(s)$.
In the next step, one identifies the divergences and subdivergencies of the 
graphs, and combines each of them with the appropriate terms of
the counterterm contributions at the level of the integrands.
In this way one renders each of the graphs ultraviolet finite.
One can then apply the numerical methods of ref. \cite{3loop} for calculating
the ultraviolet finite graphs.

At this point it becomes clear that identifying the divergences of the
multiloop graphs, choosing an appropriate definition of
the counterterms, imposing convenient renormalization conditions,
and performing the actual numerical integration are in fact closely
related issues. This is rather involved, and we will not enter into 
such details, which are
beyond the scope of this article. We will discuss this in detail 
in a subsequent publication. Here we only give a sketch of the
treatment for one of the graphs.

\begin{figure}
\begin{tabular}{cccc}
      $i A_1(s) =$
    &        \raisebox{-.99cm}{\epsfxsize = 4.4cm \epsffile{a6.eps}}  
    & \parbox{2cm}{ \[ \equiv ~~ \sum_{k = 0}^{\infty}
                                 \sum_{l = 0}^{\infty} \] }
    & \raisebox{-1.81cm}{\epsfxsize = 8cm \epsffile{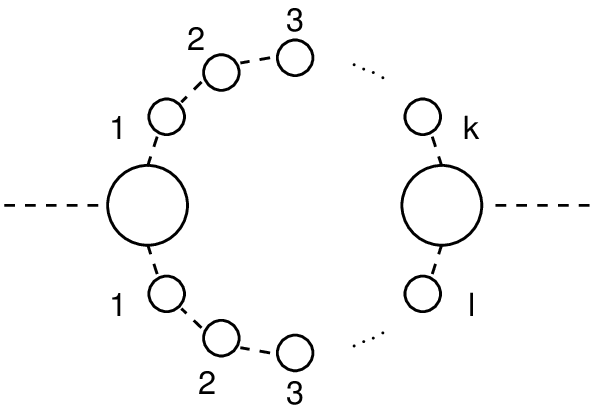}}  
\end{tabular}
\caption{{\em Multiloop diagrams with three--loop topology which
              contribute to the $\chi \chi$ propagator 
	      in next--to--leading order.}}
\end{figure}

Let us consider the graph $A_1$ which is shown in fig. 3. 
This graph has the topology of a three--loop Feynman diagram,
which is the most complicated topology among the graphs of 
fig. 2. To evaluate this graph, we treat it as if it were a three--loop
diagram with two propagators replaced by some more complicated 
expressions, given by eqns. 12. We evaluate the resulting expression
according to ref. \cite{3loop}. This gives the following two--dimensional 
integral representation:

\begin{equation}
A_1(s)  =
     \int_{-\infty}^{\infty} d p_0\,\int_{0}^{\infty} d \rho\, 
        (4 \, \pi \, \rho^2)
	D_{\chi \chi}(P_0^2) 
	D_{\chi \chi}(P_k^2)\,
        C^2(P_k^2,P_0^2,k^2) \, \; \; ,
\end{equation}
where

\begin{eqnarray}
P_0^2  & = &  p_0^2 - \rho^2  
\nonumber \\ 
P_k^2  & = & p_0^2 + 2 p_0 \sqrt{k^2} + k^2 - \rho^2
\nonumber \\ 
P_{\mu}^2  & = & p_0^2 + 2 p_0 \mu + \mu^2 - \rho^2
      \; \; ,
\end{eqnarray}
and $k=\sqrt{s}$ is the external momentum of the graph. 
$\mu$ is some arbitrary subtraction scale for the subdivergencies of the
graph, which will be used in the following.
$C(P_1^2,P_2^2,P_3^2)\equiv C(0,0,0;P_1^2,P_2^2,P_3^2)$ is the 
three--point vertex diagram, for which we use an analytical relation
given in ref. \cite{3loop}. That expression is suitable for our
NLO $1/N$ calculation because it remains on the correct Riemann
sheet for the range of parameters needed to evaluate the two--fold 
integral of eq. 16. 

The expression of eq. 16 cannot be evaluated numerically directly
because it is ultraviolet divergent. First one has to combine
it with the appropriate counterterms in order to subtract its 
divergencies and subdivergencies, thus rendering this expression 
finite. After some algebra, and after introducing the notation:

\begin{equation}
  \xi = \frac{1}{16 \pi^2 P_0^2} \log{\left(\frac{s}{\mu^2}\right)} \; \; \; ,
\end{equation}
one finds the following subtracted expression:

\begin{eqnarray}
\lefteqn{A_1(s) \, = \, \int_{-\infty}^{\infty} d p_0\,\int_{0}^{\infty} d \rho\, 
                                                    (4 \, \pi \, \rho^2) \times    
     \left\{
    D_{\chi \chi}(P_0^2) \left[ D_{\chi \chi}(P_{\mu}^2) - D_{\chi \chi}(P_0^2) \right] \xi^2
   \right. } \nonumber \\ 
  &  &  \left.  ~~~~~~~~~~~~~
+ ~ D_{\chi \chi}(P_0^2) \left[ D_{\chi \chi}(P_k^2) - D_{\chi \chi}(P_{\mu}^2) \right] C(P_k^2,P_0^2,k^2)^2
   \right.  \nonumber \\ 
  &  &  \left.
- ~ 2 D_{\chi \chi}(P_0^2) D_{\chi \chi}(P_{\mu}^2) \left[ C(P_k^2,P_0^2,k^2) - C(P_{\mu}^2,P_0^2,\mu^2) + \xi \right] \xi
   \right.  \nonumber \\ 
  &  &  \left.
+ ~ 2 D_{\chi \chi}(P_0^2) D_{\chi \chi}(P_{\mu}^2) \left[ C(P_k^2,P_0^2,k^2) - C(P_{\mu}^2,P_0^2,\mu^2) + \xi \right] C(P_{\mu}^2,P_0^2,\mu^2) 
   \right.  \nonumber \\ 
  &  &  \left.
~~ + ~ D_{\chi \chi}(P_0^2) D_{\chi \chi}(P_{\mu}^2) \left[ C(P_k^2,P_0^2,k^2) - C(P_{\mu}^2,P_0^2,\mu^2) + \xi \right]^2
   \right\}
 \; \; \; .
\end{eqnarray}

Here the subtraction scale of the subdivergencies, $\mu$, 
can be set to zero without encountering any infrared difficulties.
This choice is in fact more convenient for the actual calculation.
The reader can easily convince himself that the expression above is 
ultraviolet convergent. The expression can be evaluated
numerically fast and precisely by choosing an appropriate complex
integration path, as explained in ref. \cite{3loop}.

The other graphs can be evaluated along the same lines. For the
graphs $A_4$ and $A_5$ we found it advantageous to rewrite 
the integrands in a special
form, which results in a faster algorithm. These methods will 
be discussed in detail in future publications.

At this point we would like to make some comments related to the renormalization
procedure. First, we emphasize again that we use a renormalization
scheme at intermediary stages, and a renormalization scale $\mu$ appears in 
relations like eq. 19. However, the final physical results are totally 
independent of this renormalization scheme, just like the leading order results.
This is because the perturbation series of the $1/N$ coefficient is summed
up exactly, to all orders.

An interesting effect is that 
the wave function renormalization constants of the Goldstone and Higgs 
bosons become finite when calculated in the nonperturbative $1/N$ expansion.
This only shows up nontrivially in the NL order, as there are no contributions
to the wave function renormalization constants in leading order.
The reader can easily convince himself that the only graphs potentially
giving ultraviolet divergent contributions to the wave function renormalization
constants are $B_1$ and $D$. If one considers the Feynman diagrams which
are contained in both $B_1$ and $D$, one notices that they are all 
indeed ultraviolet divergent. However, after summing up all Feynman
diagrams which compose these two graphs, the ultraviolet divergency
of the wave functions disappears. One is left with very slowly 
(logarithmical) converging expressions.

The finiteness of wave function renormalization constants is known to be 
a feature of the exact solution for interacting fields.
In fact, one can prove they ought to be numbers with
values between 0 and 1. For the truncated perturbative solution this is
notoriously not the case, wave function renormalization constants being
even ultraviolet divergent in general. For the case of the NLO calculation
described here, we were able to prove that the Higgs and Goldstone wave
function renormalization constants are finite. However, we could not
so far check that their values are between 0 and 1. It is difficult
to calculate their actual values numerically because their numerical 
convergence is only logarithmical and much slower than that of the
graphs needed for physical calculations. 

The behaviour of higher orders in $1/N$ is an open question.
Still, it is remarkable that the NLO of the $1/N$ expansion recovers 
such a property of the exact solution which is absent in perturbation theory.


\section{Nonperturbative Higgs lineshape}

The Higgs propagator which we calculated in the previous section
is essentially a physical quantity. It can in principle be measured in a
fermion scattering process, for instance the 
$\mu \bar \mu \rightarrow t \bar t$ process, which is relevant for muon collider studies. The only contribution which one still has to add is the 
correction to the two Yukawa couplings of the Higgs to the fermions.
As long as one only considers the contributions of leading order
in the mass of the fermions, it is well--known \cite{2loop,riess,jikia} 
that these contributions
are energy independent, and are simply given by the ratio of the wave
function renormalization constants of the Higgs and the Goldstone bosons.   
Regarding the evaluation of this ratio, we already mentioned in the
previous section that $Z_{\sigma}$ and $Z_{\pi}$ are both finite at NLO
in $1/N$,
but evaluating them separately is very difficult because they 
converge very slowly. However, they have the same ultraviolet behaviour
and for this reason their ratio is well--behaved numerically. In this
context one can note that this ratio is ultraviolet finite even 
in perturbation theory.

The results for the lineshapes are given in fig. 4.
We plot the lineshape function 
$f(\sqrt{s}) = | D_{\sigma \sigma}(s) Z^2_{\sigma \sigma}/Z^2_{\pi \pi} |^2$,
both at LO and at NLO in the $1/N$ expansion.
We also compare with the known perturbative results 
up to NNLO \cite{binoth}. 
For the purpose of comparing the predictions of perturbation theory
and of the $1/N$ expansion, we compare in fig. 4 Higgs lineshapes
which have the same position of the peak.

\begin{figure}
\hspace{1.5cm}
    \epsfxsize = 15cm
    \epsffile{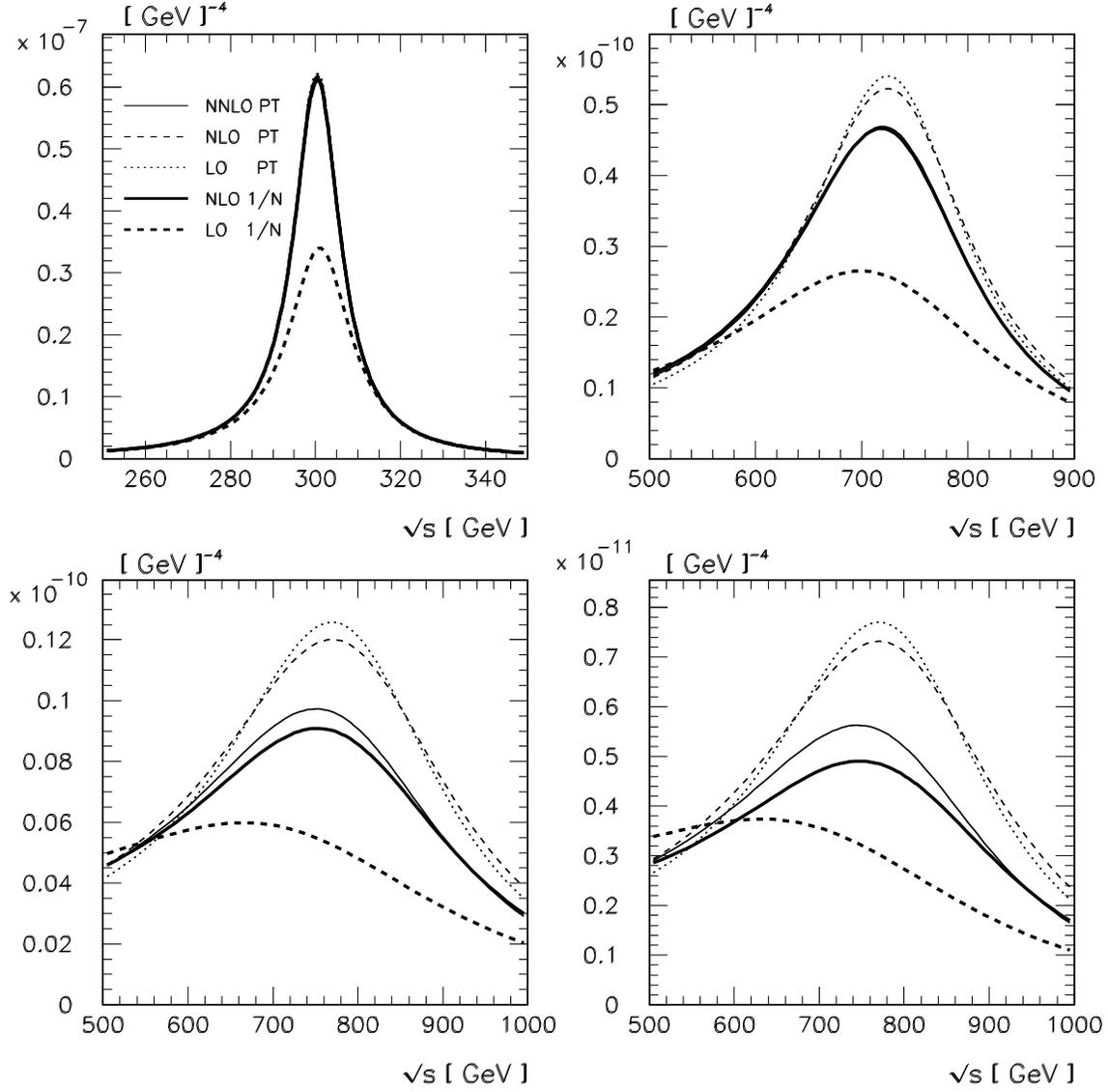}
\caption{{\em The Higgs lineshape in fermion scattering. Comparison
              between the perturbative lineshape (LO, NLO, NNLO) and
	      the nonperturbative $1/N$ expansion (LO, NLO) 
	      for different values of the coupling. 
	      We compare lineshapes so that the position 
	      of the peak in two--loop perturbation theory and
	      NLO $1/N$ expansion is the same. The lineshapes in 
	      NNLO perturbation theory and NLO $1/N$ are almost
	      indistinguishable up to about 700 GeV.}}
\end{figure}

As one can see, there is a substantial discrepancy between the $1/N$ expansion
in leading order and perturbation theory even for low values of the coupling.
We notice that our NLO result lies between the perturbative results and the 
lowest order in $1/N$. The corrections are actually quite large compared to the
lowest order in $1/N$. For low coupling we see that one is very close to
perturbation theory now, which indicates that we should have 
a good approximation for the Higgs line shape. For Higgs 
masses larger than about 850 GeV there is a 
fairly large difference between perturbation 
theory and the 1/N expansion. Perturbation theory is probably not 
reliable anymore, but things seem to be converging. 

\begin{figure}
\hspace{1.5cm}
    \epsfxsize = 15cm
    \epsffile{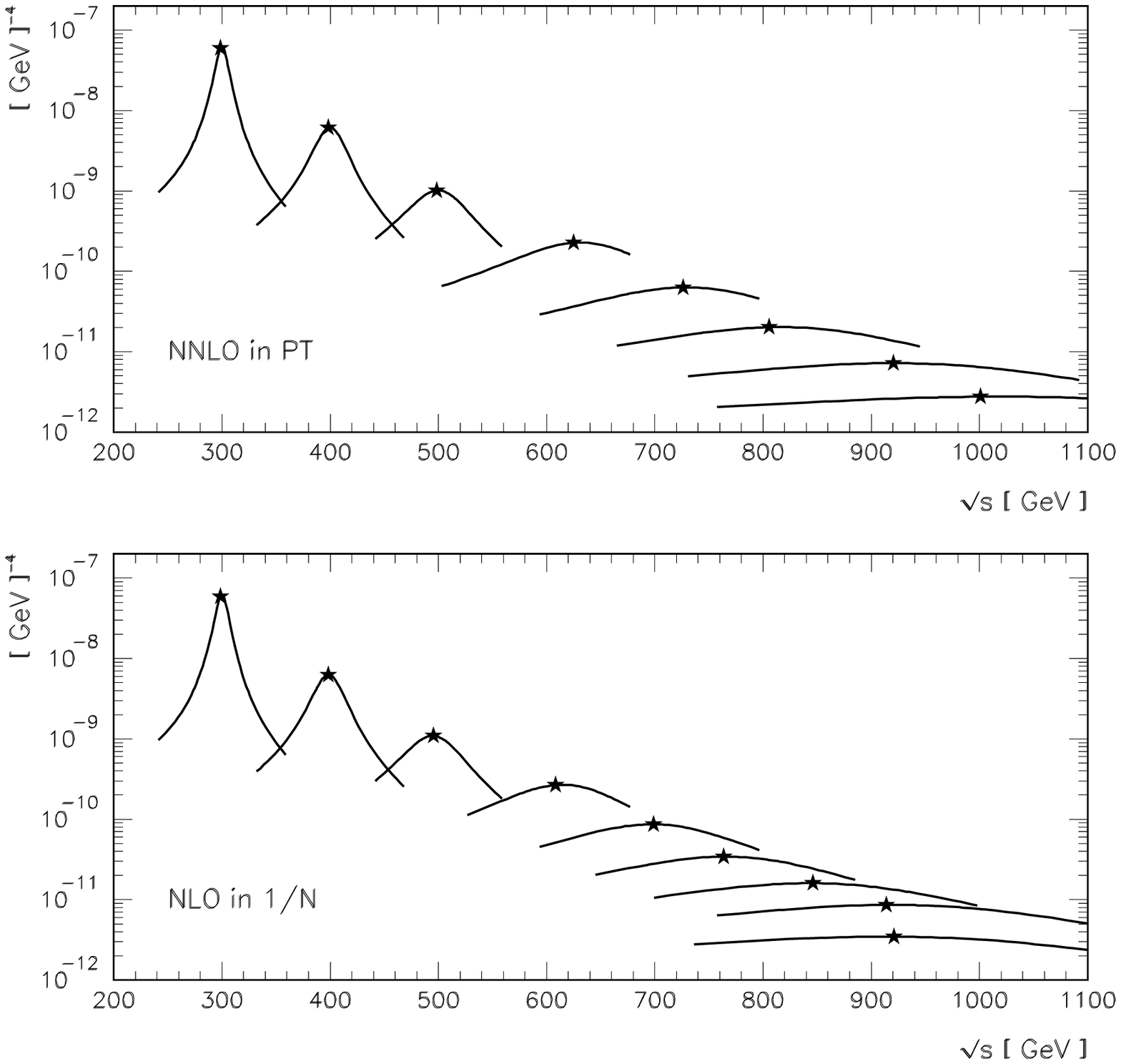}
\caption{{\em Higgs lineshapes in perturbation theory (two--loop) 
              and in the nonperturbative $1/N$ expansion (NLO),
	      where we marked the position of the peak.
	      One can see the saturation of the position of the peak
	      for large values of the coupling.}}
\end{figure}

For even larger values of the coupling the behaviour of the $1/N$
solution and of perturbation theory becomes different, and lineshapes
cannot be compared at the same position of the peak any longer.
This behaviour can be seen in fig. 5. Here we plot lineshapes in
NLO $1/N$ expansion and in two--loop perturbation theory for a range 
of couplings. Since perturbation theory breaks down for large couplings,
the perturbative plots for heavy Higgs bosons are given only 
for comparison purposes. While the position of the peak in perturbation theory 
increases continuously with the coupling, the $1/N$ solution shows
a saturation around 930 GeV, after which only the decay width continues
to increase.

\begin{figure}
\hspace{2.5cm}
    \epsfxsize = 14.5cm
    \epsffile{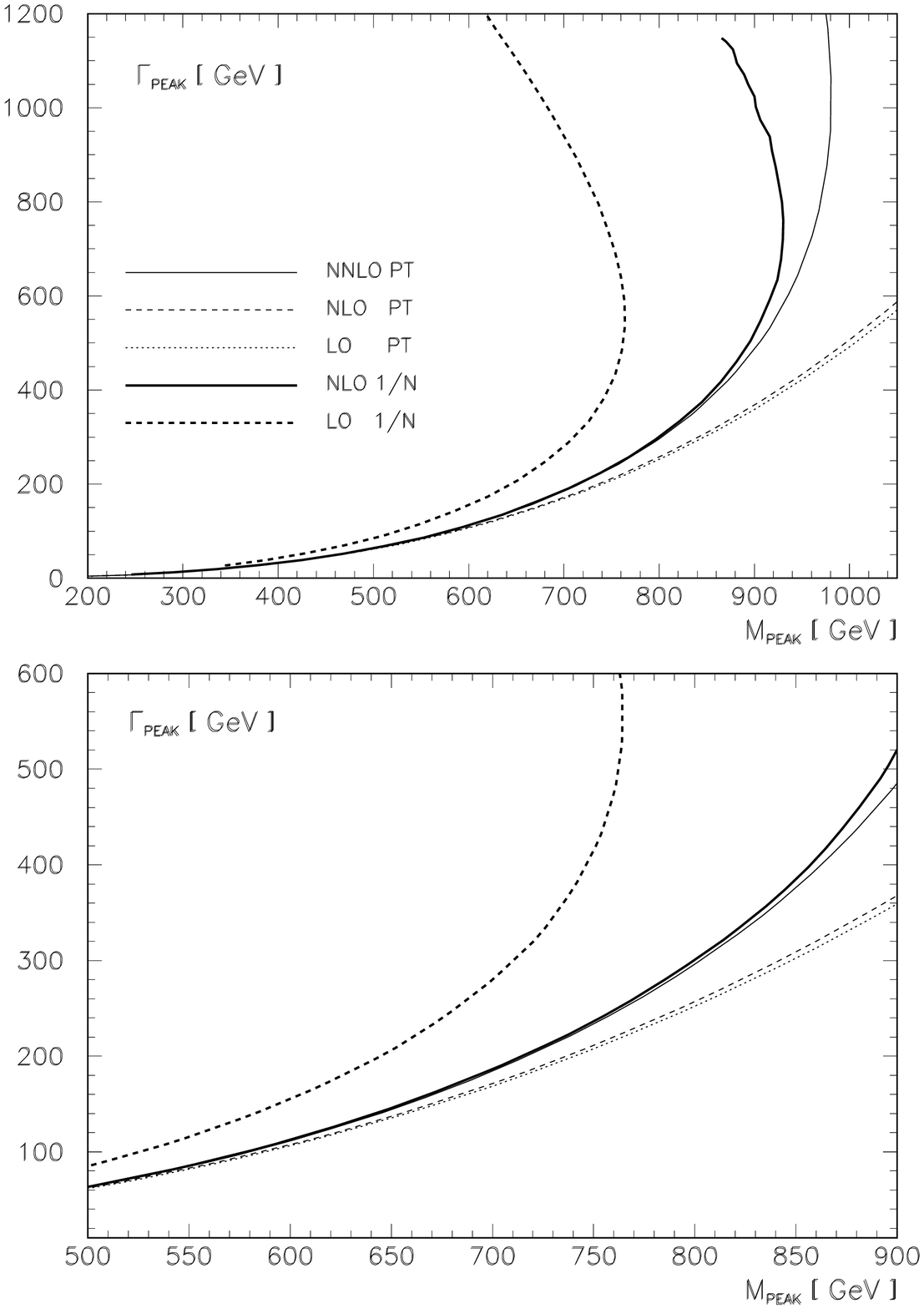}
\caption{{\em The relation between the peak variables $M_{PEAK}$ 
              and $\Gamma_{PEAK}$
              in perturbation theory and in the $1/N$ expansion.}}
\end{figure}

In order to study this in somewhat more detail, we plotted in fig. 6
two observable quantities which can be interpreted as an
effective Higgs mass and width. Here, ideally one would like to plot instead
the real and the imaginary part of the pole of the Higgs propagator because
they are universal for all Green functions and are not process dependent.
However our numerical programs do not allow us at the moment to calculate 
the graphs on the second Riemann sheet and solve the pole equation.

Therefore we use a procedure to define an effective mass $M_{peak}$
and width $\Gamma_{peak}$ from the lineshape itself, which is
more directly related to experiment. The mass $M_{peak}$
is defined as the location of the peak of the cross section.
The width $\Gamma_{peak}$ is taken from the peak height of the
cross section as if the lineshape was determined by a resonant 
Higgs propagator only, giving rise to a Breit--Wigner shape:

\begin{equation}
  \sigma(s) \sim \frac{1}{(s - M_{peak})^2
                             + M_{peak}^2 \Gamma_{peak}^2}
  ~~~~~~ .
\end{equation}

Of course, the lineshape is not exactly a Breit--Wigner resonance. 
However, we found that this definition of the width is within 
a few percent of for instance the width at half--maximum. 
The description by $M_{peak}$ and $\Gamma_{peak}$ gives therefore 
a satisfactory approximation of the leading features of the Higgs 
lineshape. It should be good enough to be used as a first 
approximation for phenomenological applications.

Our results can strictly speaking not be used directly at the LHC.
This is because one should also take higher order $1/N$ 
correction in the Higgs-Goldstone boson vertex into account. 
However, we expect that the 
correction to the width will be the dominant feature in practice.
For values of the Higgs mass larger than about 900 GeV, 
the saturation effect sets in, and 
the uncertainty due to the discrepancy between the perturbative 
and the $1/N$ widths becomes important. The Higgs
becomes wider, and the signal is washed out and lost in the background.
This saturation effect can be quite important for LHC physics 
and the discovery limits there.    						     
	
We plot in fig. 6 the effective width $\Gamma_{peak}$ as a function of
$M_{peak}$ both in the $1/N$ expansion and in perturbation theory.
One can see from fig. 6 that the $1/N$ expansion and perturbation theory appear 
to be converging towards a common relation between the Higgs width and mass. 
For a Higgs mass up to about 850 GeV, the agreement is remarkable. 
Afterwards the numerical values start to deviate. 
The nonperturbative width is larger 
than the width in perturbation theory, and saturation sets in. 
A maximum of the effective Higgs mass appears at around 930 GeV. 

For phenomenological purposes we give in the following a simple approximate
formula to relate $M_{peak}$ and $\Gamma_{peak}$.
Expressing $M_{peak}$ and $\Gamma_{peak}$ in TeV, one can use the 
following relation:

\begin{eqnarray}
  M_{peak}^3 & = &   2.02582  \cdot \Gamma_{peak}   
	           - 0.812725 \cdot \Gamma_{peak}^2 
   \nonumber \\
      	    & & 
                   - 1.01729  \cdot \Gamma_{peak}^3 
	           + 0.541511 \cdot \Gamma_{peak}^4  
		~~~~ ,
\end{eqnarray}
which is simply a fit of the NLO $1/N$ curve shown in fig. 6. Its precision
is at the per mille level for the range of couplings shown in fig. 6.
For comparison, the corresponding fit for the two--loop perturbative 
curve is the following (the exact two--loop lineshape is given 
in ref. \cite{binoth}):

\begin{eqnarray}
  M_{peak}^3 & = &   2.09069  \cdot \Gamma_{peak}   
	           - 1.23887 \cdot \Gamma_{peak}^2 
   \nonumber \\
      	    & & 
                   - 0.0233177  \cdot \Gamma_{peak}^3 
	           + 0.115571 \cdot \Gamma_{peak}^4  
		~~~~ .
\end{eqnarray}


\section{Conclusions}

We treated an $O(N)$--symmetrical sigma model with spontaneous
symmetry breaking in the nonperturbative $1/N$ expansion at
next--to--leading order.

In order to calculate corrections of higher order in $1/N$
one needs a prescription for the treatment of the tachyons which appear
in the propagators at leading order. We showed that in the usual derivation
of the $1/N$ expansion the occurrence of tachyons is arbitrary, and
it can be eliminated when one realizes that the coefficient of the $1/N$ expansion
is only determined up to a function of the coupling constant which vanishes
identically in perturbation theory. Therefore we introduced a scheme
of tachyonic regularization which can be performed consistently order
by order in the $1/N$ expansion. Our tachyonic regularization can
be interpreted as a prescription for the summation of the coefficient 
of the $1/N$ expansion which leaves unchanged its perturbative expansion
and preserves causality at the same time.

The tachyonic regularization is not unique. This arbitrariness can be used
for modelling nondecoupling effects of unknown physics at a higher energy
scale. This arbitrariness is an interesting problem which deserves further 
investigation.

We then developed techniques to calculate infinite sets of multiloop 
Feynman diagrams which are needed for higher order calculations in $1/N$.
We applied these techniques to the Higgs sector of the standard model,
$N=4$, and calculated the Higgs propagator and the lineshape of the Higgs
resonance at muon colliders. The results are interesting, and confirm
qualitatively some results obtained previously based on the leading order 
solution, such as the saturation of the Higgs mass. In NLO, it
occurs at a value of about 930 GeV.
Moreover, compared to the leading order, the NLO correction is rather substantial.
It leads for low couplings to a remarkable agreement between perturbation 
theory and $1/N$ expansion, thus solving a long--standing puzzle.
This agreement proves that the nonperturbative $1/N$ expansion in higher
orders, when combined with the tachyonic regularization, is a useful
tool for treating strong couplings. It can give very precise results.

Our NLO solution has certain properties associated with a nonperturbative 
solution. For instance, it contains no residual renormalization scheme
dependence. Of course, this is true for the leading order as well. An 
interesting property which only shows up at NLO is the finiteness
of wave function renormalization constants, which also ought to be satisfied
by the exact solution, and which does not hold in perturbation theory.

Regarding Higgs searces at future colliders, it is interesting that the
perturbative two--loop correction is substantial when
compared to the one--loop and tree level, 
but turns out to be close to the nonperturbative solution
which we derived up to about 900 GeV. This reduces
drastically the theoretical uncertainty of a heavy Higgs width.

For larger couplings, our results show a deviation from perturbation
theory. Essentially, the Higgs width is always larger than in perturbation
theory. The Higgs mass is saturated, and its width can grow without its mass
becoming larger at the same time. This effect must be taken into account
in view of heavy Higgs searches at LHC. For phenomenological purposes 
we gave a simple approximate formula to relate the Higgs mass and width 
which is valid for large couplings as well.



\vspace{.5cm}

{\bf Acknowledgements}

We would like to thank George Jikia and Boris Kastening for discussions.
The work of A. G. was supported by the Deutsche Forschungsgemeinschaft (DFG).


\newpage


\end{document}